\documentclass[floatfix,twocolumn,prl]{revtex4}
\usepackage{graphicx}
\usepackage{amsmath}
\usepackage{amssymb}
\usepackage{bm}
\usepackage{dcolumn}
\usepackage{braket}

\usepackage{color}

\begin{document}

\title{The rise of spin-flip transitions in the anomalous Hall effect of FePt alloy}

\author{Hongbin~Zhang}
\author{Frank Freimuth}
\author{Stefan~Bl\"ugel}
\author{Yuriy~Mokrousov}
\email[corresp.\ author: ]{y.mokrousov@fz-juelich.de}
\affiliation{Institut f\"ur Festk\"orperforschung and Institute for Advanced Simulation, 
Forschungszentrum J\"ulich and JARA, D-52425 J\"ulich, Germany}
\author{Ivo~Souza}
\affiliation{Department of Physics, University of California, Berkeley, CA 94720, USA}
\date{\today}

\def\sigpar{\sigma_m}
\def\sigperp{\sigma_\theta}
\def\sigout{\sigma_z}
\def\sigin{\sigma_x}
\def\alphapt{\alpha^{\rm Pt}}
\def\alphapd{\alpha^{\rm Pd}}
\def\alphafe{\alpha^{\rm Fe}}
\def\uu{\upuparrows}
\def\ud{\uparrow\hspace{-0.055cm}\downarrow}

\def\red#1{{\color{red}#1}}

\begin{abstract}

  We carry out {\it ab initio} calculations which demonstrate the
  importance of spin-flip transitions for the intrinsic anomalous Hall
  conductivity of ordered FePt alloys.  We show the such transitions 
  get enhanced by large spin-orbit coupling of Pt atoms, becoming 
  negligible when Pt is replaced by lighter isoelectronic Pd. We find 
  that spin-flip transitions in FePt originate not only from conventional 
  band anticrossings at the Fermi level, but also from transitions 
  between well-separated pairs of bands with similar dispersions. 
  We also predict a strong anisotropy in the anomalous Hall 
  conductivity of FePt, which comes from spin-flip transitions 
  entirely, and investigate the influence of disorder on it. 

\end{abstract}

\maketitle

The intrinsic anomalous Hall effect (AHE)~\cite{Nagaosa:review} and
  spin Hall effect (SHE)~\cite{Hirsch} in solids arise from the opposite
  anomalous velocities experienced by spin-up and spin-down electrons as
  they move through the spin-orbit-coupled bands under an applied electric 
  field. In paramagnets, where the bands are normally spin-degenerate, 
  these counter-propagating transverse currents result in a time-reversal 
  ($T$) conserving pure spin current. In ferromagnets, where the bands 
  are split by the exchange interaction, the same process generates a net 
  $T$-odd charge current.

The above picture is intuitively appealing, and often leads to correct
  conclusions.  However, it leaves out the fact that in the presence of
  spin-orbit coupling the spin projection along the quantization axis is
  not a good quantum number. This is a particularly subtle point regarding 
  the SHE, as the proper definition of the spin current becomes problematic 
  when spin is not a conserved quantity~\cite{Shi}. More generally, 
  processes which do not conserve  spin are known to play a role in 
  phenomena such as spin relaxation~\cite{Zutic} and magnetocrystalline
  anisotropy~\cite{MAE}. It is however usually assumed that such 
  {\it spin-flip} processes can be safely ignored when studying
  transport. This viewpoint is supported by recent calculations of the
  intrinsic AHE~\cite{Kontani} and extrinsic SHE~\cite{Gradhand}.

In this work we use first-principles calculations to study the impact 
  of spin-flip transitions on the intrinsic anomalous Hall conductivity 
  (AHC) of FePt orderered alloys~\cite{Seemann}. We find that their 
  effect is considerable, as they account for about one fifth of the net 
  AHC. More importantly, the calculations reveal a clear experimental 
  signature of the spin-flip AHC: as the magnetization is rotated from 
  the uniaxial direction to the basal plane, the spin-flip contribution 
  changes sign, leading to a factor-of-two reduction in the net AHC, 
  while the spin-conserving part is almost perfectly isotropic.

We identify two distinct mechanisms for the spin-flip transitions. The 
  first involves spin-orbit-induced anticrossings at the intersections of 
  up- and down-spin Fermi-surface sheets (the locus of intersection 
  forms loops in $k$-space, which we shall refer to as {\it hot loops}, 
  in analogy with the {\it hot spots} that have been discussed in 
  connection with spin-relaxation~\cite{Zutic}). The second mechanism 
  involves spin-orbit driven transitions between bands with similar 
  dispersion which are split in energy across the Fermi level. We shall 
  refer to them as {\it ladder} transitions. Both features occur at very 
  low frequencies, of the order of the spin-orbit coupling strength.

Let us briefly review the formalism for calculating the intrinsic AHC
  from first-principles.  For a ferromagnet with the orthorhombic
  crystal structure and magnetization $\mathbf{M}$ along the $\hat{z}$
  ([001]) axis, the intrinsic anomalous Hall conductivity (AHC) 
  $\sigma_{z}\equiv\sigma_{xy}$ is given by the $k$-space integral 
  of the Berry curvature~\cite{Yao,Nagaosa:review}:
  \begin{equation}
   \sigma_{z} = \frac{e^2\hbar}{4\pi^3}{\rm Im}\int_{\rm BZ} 
   d\mathbf{k} \sum_{n,m}^{o,e}
   \frac{\Braket{\psi_{n\mathbf{k}}|v_x|\psi_{m\mathbf{k}}}
   \Braket{\psi_{m\mathbf{k}}|v_y|\psi_{n\mathbf{k}}}}
   {(\varepsilon_{m\mathbf{k}}-\varepsilon_{n\mathbf{k}})^2}.
  \end{equation}
  In this expression $\psi_{n\mathbf{k}}$ and $\psi_{m\mathbf{k}}$ 
  are respectively the occupied ($o$) and empty ($e$) spinor Bloch
  eigenstates of the crystal, $v_x$ and $v_y$ are components of the
  velocity operator $\mathbf{v}$, and the integral is over the Brillouin
  zone (BZ).  When the direction of $\mathbf{M}$ is changed from the 
  $\hat{z}$-axis to the $\hat{x}$-axis ([100]), the 
  $\sigma_{x}\equiv\sigma_{yz}$ component of the conductivity 
  tensor should be calculated instead, by replacing $v_x\rightarrow v_y$ 
  and $v_y\rightarrow v_z$ in Eq.~(1).

The calculations were done using the approach of Ref.~\cite{Souza:2006}, 
  whereby the linear-response expression (1) is rewritten in the basis of
  Wannier functions spanning the occupied and low-lying empty states. 
  In this way the infinite sums over bands are replaced by sums over the 
  small number of Wannier-interpolated bands.  The Wannier functions 
  were generated with {\tt WANNIER90}~\cite{wannier} using the same 
  parameters as in Ref.~\cite{Seemann}, by post-processing first-principles 
  calculations done using the J\"ulich DFT FLAPW code {\tt FLEUR}~\cite{fleur} 
  (see Ref.~\cite{FLAPW-MLWFs} for details).  The unit cell contained two 
  atoms in the $L1_0$ structure, with stacking along the [001]-direction 
  and lattice constants $a=5.14$~a.u.~and $c=7.15$~a.u.~\cite{footnote2}.

The spin-orbit term in the Hamiltonian has the form
  \begin{equation}
  \label{eq:soi}
    \xi\mathbf{L\cdot S}=
    \xi{\rm L}_{\hat{n}}{\rm S}_{\hat{n}} + \frac{\xi}{2}\left( 
    {\rm L}^{+}_{\hat{n}}{\rm S}^{-}_{\hat{n}} + 
    {\rm L}^{-}_{\hat{n}}{\rm S}^{+}_{\hat{n}} \right),
  \end{equation}
  where $\xi$ is the spin-orbit coupling strength and $\hat{n}$ is the
  magnetization direction, which is taken as the spin-quantization
  axis. The first term on the~r.h.s.~of Eq.~(\ref{eq:soi}), which we
  will denote as ${\rm LS}^{\uu}$, preserves the spin of a pure-spin
  state $\psi_{n\mathbf{k}}$, while the second term, ${\rm LS}^{\ud}$, 
  flips it. The spin-flip ($\sigma^{\ud}$) and spin-conserving ($\sigma^{\uu}$) 
  parts of the AHC $\sigma$ are calculated from Eq.~(1) after selectively turning 
  off the ${\rm LS}^{\ud}$ or ${\rm LS}^{\uu}$ parts of the SOI
  Hamiltonian~(\ref{eq:soi}).  We find that to a very good approximation
  they are additive, i.e., $\sigma \approx
  \sigma^{\uu}+\sigma^{\ud}$.  

\begin{table}[t!]
\begin{ruledtabular}
\begin{tabular}{ccccccc}
    &   & Total & $\sigma^{\uu}$ & $\sigma^{\ud}$ & $\Delta\sigma^{\uu}$ 
    & $\Delta\sigma^{\ud}$ \\ \hline
    FePt & $[001]$ & 818.1 &  576.6 & 133.4    &  $-$8.5  &  317.3  \\
         & $[100]$ & 409.5 &  585.1 & $-$183.9 &          &         \\ \hline
    FePd  & $[001]$ & 135.1 &  108.4 &  28.4    & $-$88.5  & $-$33.6 \\
          & $[100]$ & 275.9 &  196.9 &  62.0    &          &         \\
\end{tabular}
\end{ruledtabular}
\label{table:anisotropy}
\caption{
  Values of the AHC for [001] ($\sigout$) and [100] ($\sigin$) directions of magnetization
  $\mathbf{M}$ in FePt and FePd. $\Delta\sigma^{\uu(\ud)}$ 
  is defined as the difference between no-flip (flip) parts of $\sigma_z$ and
  $\sigma_x$. All values are in S/cm.}
\end{table}

The importance of spin-flip transitions for the AHC of FePt can be seen 
  by analyzing its dependence on the magnetization direction (see Table~1).  
  If only the spin-conserving term in Eq.~(\ref{eq:soi}) is kept, the resulting 
  AHC $\sigma^{\uu}$ changes by less than 2\% from an average value of 
  about 580~S/cm as ${\bf M}$ is tilted from the $\hat{z}$-axis to the 
  $\hat{x}$-axis.  When the the spin-flip term is also included in the calculation, 
  the AHC becomes highly anisotropic, decreasing by a factor of two from [001]
  to [100]. Keeping only the spin-flip part of the SOI reveals that it is indeed 
  responsible for the large anisotropy, as the resulting AHC changes by more 
  than 300~S/cm, from a positive value along [001] to a negative value along 
  [100]. Such large AHC anisotropy can occur in uniaxial crystals, and was 
  previously found in hcp Co~\cite{Souza:2009}, however, as opposed to FePt,
  in hcp Co the anisotropy is caused for the most part by spin-conserving 
  transitions.
 
Such a large spin-flip contribution in FePt is rather unexpected, in view of the 
  fact that in a perturbative expansion in powers of $\xi$ only the spin-conserving 
  part of the SOI Hamiltonian (\ref{eq:soi}) contributes to Eq.~(1), with spin-flip
  appearing only at second order~\cite{Cooper}. It should however be kept in 
  mind that because Pt is a heavy atom, the SOI cannot be treated as a small 
  perturbation in FePt. Moreover, the AHC is very sensitive to near-degeneracies 
  across the Fermi level~\cite{Yao}, and therefore the above analysis of 
  Ref.~\cite{Cooper}, which is based on non-degenerate perturbation theory, 
  may not apply. 

\begin{figure}[t]
\begin{center}
\includegraphics[scale=0.45]{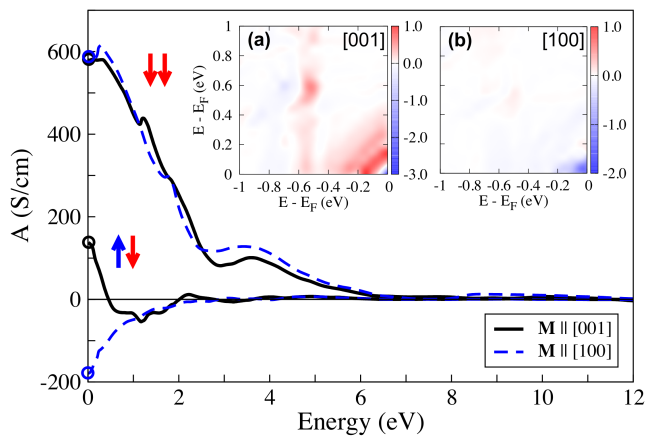}
\end{center}
\caption{\label{fig:cumulative-ahc} (color online) Spin-flip ($\ud$)
  and spin-conserving ($\uu$) cumulative contribution to the AHC from
  the spectrum above energy $\omega$, $A(\omega)$. Inset presents the
  energy-energy density of contributions to the flip-AHC,
  $\Sigma(\varepsilon_1,\varepsilon_2)$, for $\mathbf{M}$ along (a)
  [001] and (b) [100]-axes (in $10^5$~a.u./${\rm eV}^2$).
  Open circles indicate the values of AHC from Table 1.}
\end{figure}

The AHC can be resolved in energy by defining a {\it cumulative} AHC 
  $A(\omega)$, which accumulates all transitions in Eq.~(1) for which 
  $\varepsilon_{m\mathbf{k}}-\varepsilon_{n\mathbf{k}}>\omega$~\cite{Souza:2009}.
  In the limit $\omega\rightarrow 0$ all interband transitions are accounted for, 
  and $A(\omega=0)$ equals the full AHC. The spin-conserving and spin-flip 
  contributions to the cumulative AHC are plotted in Fig.~\ref{fig:cumulative-ahc} 
  in the range $0\leq\omega\leq 12$~eV, for both $\mathbf{M}\Vert\hat{z}$ 
  and $\mathbf{M}\Vert\hat{x}$. While $A^{\uu}(\omega)$ remains largely 
  isotropic over the entire energy range and decays rather slowly with $\omega$ up 
  to 4$-$5~eV in energy, the $A^{\ud}(\omega)$ contribution picks up only for 
  $\omega$ below 1~eV and immediately becomes strongly anisotropic with 
  decreasing energy, displaying a characteristic bifurcation shape~\cite{Souza:2009}. 
  Thus, the anisotropy in the AHC arises from spin-flip transitions
  in the 0.5~eV~energy window around $E_F$. 

To get a further insigt into which kind of transitions is responsible for 
  $A^{\ud}(\omega)$, we calculate the density of contributions to the AHC 
  given by Eq.~(1), 
  $\Sigma(\varepsilon_1,\varepsilon_2)$, from the states with energies 
  $\varepsilon_1<E_F$ and $\varepsilon_2>E_F$. The overall integral 
  $\iint\Sigma(\varepsilon_1,\varepsilon_2) d\varepsilon_1d\varepsilon_2$ 
  provides the value of $\sigma^{\ud}$,  while assuming the constraint 
  $\varepsilon_2-\varepsilon_1>\omega$, this integral gives the value of 
  $A^{\ud}(\omega)$. Density $\Sigma$ can be used in combination 
  with the cumulative AHC to obtain more information on the energy structure 
  of the Berry curvature and the anomalous Hall conductivity.

The calculated density $\Sigma$ for $\mathbf{M}\Vert\hat{z}$ and
  $\mathbf{M}\Vert\hat{x}$ is presented in Fig.~1(a) and (b), respectively.
  In these plots we can clearly see the contributions from the band 
  anticrossings of $\uparrow$- and $\downarrow$-bands along the hot loops
  in the BZ. They are given by blue dots around the origin 
  $\varepsilon_1=\varepsilon_2=0$ and provide a negative contribution to the
  AHC for both magnetization directions. While for $\mathbf{M}\Vert\hat{x}$
  the hot loops contribution dominates, for $\mathbf{M}\Vert\hat{z}$
  a competing positive contribution to the AHC can be clearly seen in
  Fig.~1(a). It is given by series of stripes 
  $\varepsilon_2-\varepsilon_1\approx\rm{const.}$ in the vicinity of the Fermi 
  energy. By analyzing the band structure we find that these transitions
  come from pairs of bands of different orbital character with
  similar dispersion around $E_F$ (see inset in Fig.~2). Such ladder
  transitions, induced by SOI, provide a different source of the AHC as
  they do not require a band crossing at the Fermi energy, and occur
  over large regions in energy and $k$-space. In case of FePt with 
  $\mathbf{M}\Vert\hat{z}$ their contribution is so large that it wins 
  over the hot-loops part and determines the sign and magnitude of 
  the flip-AHC. 

 \begin{table}[t!]
 \begin{ruledtabular}
 \begin{tabular}{ccccccc}
     & ${\rm Fe}^{\rm tot}$  & ${\rm Fe}^{\uu}$ &  ${\rm Fe}^{\ud}$ &
       ${\rm Pt}^{\rm tot}$  & ${\rm Pt}^{\uu}$ &  ${\rm Pt}^{\ud}$ \\ \hline
     $[001]$ & $-$13.7    & 17.9 &  $-$26.8  &  848.0      &  541.0  &  282.3  \\
     $[100]$ &   210.0    & 253.6 & $-$37.5  &  65.0      &  425.7  & $-$360.6   \\
 \end{tabular}
 \end{ruledtabular}
 \label{Table3}
 \caption{
  Atomically resolved values of the AHC for [001] and [100] directions of
  magnetization $\mathbf{M}$ in FePt, decomposed into $\uu$- and $\ud$-contributions. 
  All values are in S/cm.}
 \end{table}

The spin-flip contribution to the AHC in FePt originates from heavy Pt atoms. 
  To demonstrate this we decompose the spin-orbit part of the Hamiltonian 
  in FePt in real space as
  \begin{equation}
    H^{\rm SO} =  \xi_{\rm Fe}\mathbf{L}^{\rm Fe}\cdot\mathbf{S}+
    \xi_{\rm Pt}\mathbf{L}^{\rm Pt}\cdot\mathbf{S},
  \end{equation}
  where $\mathbf{L}^{\mu}$ is the orbital angular momentum operator
  associated with atom $\mu$ and $\xi_{\mu}$ is the averaged over valence 
  $d$-orbitals spin-orbit coupling strength.
   For an Fe atom in FePt the SOI strength $\xi_{\rm Fe}^0$ amounts 
   to 0.06~eV, with corresponding value of $\xi_{\rm Pt}^0=0.54$~eV.

By using together representations of SOI according to~(2)~and~(3), we
  perform a $\uu$- and $\ud$-decomposition of the AHC coming separately
  from Fe and Pt atoms, presenting 
  results in Table~2. It can be seen, that when we consider contribution to 
  the AHC from only Fe atoms by setting $\xi_{\rm Pt}$ in (3) to zero,
  for $\mathbf{M}\Vert\hat{z}$ both $\uu$- and $\ud$-AHC are very small, 
  and while the Fe-driven $\sigma^{\ud}$ remains also very small for
  the in-plane magnetization, $\sigma^{\uu}$ dominates in this case. On 
  the other hand, by analyzing the Pt-originated AHC ($\xi_{\rm Fe}=0$), 
  we observe that the spin-flip part is very large for both magnetization
  directions. For $\mathbf{M}\Vert\hat{z}$ the sum of Pt $\sigma^{\uu}$
  and $\sigma^{\ud}$ results in a large total AHC manifesting that for this 
  magnetization direction the AHE is driven by Pt atoms. For 
  $\mathbf{M}\Vert\hat{x}$, the Pt $\sigma^{\ud}$ is of the same
  magnitude but of opposite sign to its $\uu$-counterpart, and both
  conductivities almost cancel $-$ in this case the Hall current in FePt
  is mainly of Fe origin.

 \begin{figure}[t!]
 \begin{center}
 \includegraphics[scale=0.36]{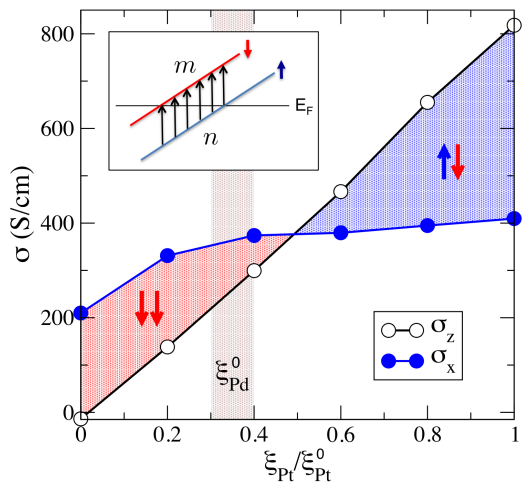}
 \end{center}
 \caption{ (color online) Dependence of $\sigout$ (open
   circles) and $\sigin$ (filled circles) in FePt alloy on the
   SOI strength inside the Pt atoms $\xi_{\rm Pt}$ with respect to its
   unscaled value $\xi_{\rm Pt}^0$. Arrows indicate the spin-flip
   ($\ud$) or spin-conserving ($\uu$) nature of the AHC
   anisotropy. Inset shows transitions in Eq.~(1) leading to the
   ladder contribution to the flip-AHC.}
\end{figure}

When decreasing the SOI strength in FePt by substituting Pt atoms with Pd 
  atoms in FePd alloy, we see an essential decrease in the values of spin-flip AHC, 
  as compared to the total AHC for both magnetization directions (Table~1), 
  which is in correspondence to the perturbation theory arguments~\cite{footnote1}.  
  For FePd, the AHC anisotropy is mainly driven by $\Delta\sigma^{\uu}$, which
  is of the same sign and somewhat larger magnitude than $\Delta\sigma^{\uu}$
  in FePt, while the total AHC anisotropy is opposite in sign to that in FePt,~c.f.~Table~1.
  In the following we make sure that such differences between 
  the AHC of FePt and FePd are indeed caused by different SOI strength and not 
  by other details of the electronic structure, such as Stoner parameters, spread 
  of $d$-functions etc.

Without affecting SOI on Fe atoms, we scale down the $\xi_{\rm Pt}^0$ 
  constant in FePt, Eq.~(3), and perform self-consistent calculations of the 
  $\sigout$ and $\sigin$ conductivities with the new corresponding SOI 
  strength, $\xi_{\rm Pt}$. The results of these calculations are presented 
  in Fig.~2 as a function of $\xi_{\rm Pt}/\xi_{\rm Pt}^0$. In this figure 
  we observe, that while the decay of $\sigout$ is almost perfectly linear, 
  the $\sigin$ conductivity stays almost constant until the regime of 
  $\xi_{\rm Pt}$ corresponding to the SOI strength of Pd atoms in FePd alloy 
  $\xi_{\rm Pd}^0=0.19$~eV (indicated with shaded area in Fig.~2). 
  This can be explained 
  by the fact that for $\mathbf{M}\Vert\hat{x}$ the AHC comes from Fe 
  atoms and thus it is not sensitive to $\xi_{\rm Pt}$, however, the AHC 
  for $\mathbf{M}\Vert\hat{z}$ is mostly Pt-originated (see Table~2), 
  with $\xi_{\rm Pt}$ serving as effective SOI strength of the system in 
  this case. 

At the value of $\xi_{\rm Pt}\approx\xi_{\rm Pt}^0/2$, $\sigin$ starts 
  dominating over $\sigout$, with values of AHC and its anisotropy 
  qualitatively close to those in FePd, when $\xi_{\rm Pt}$ reaches 
  $\xi_{\rm Pd}^0$. We conclude that the difference in value and
  sign of the AHE anisotropy between FePt and FePd alloys can indeed 
  be attributed solely to the SOI strength of Pt and Pd atoms, and the 
  crossover between $\downdownarrows$- and $\ud$-contributions with 
  the SOI strength thus explains different sign of the AHC anisotropy in 
  FePt and FePd, Fig.~2. Therefore, we suggest to use Pd-doped FePt in 
  order to tune the effective SOI strength as well as spin-character and 
  sign of the AHE anisotropy in these alloys.

 \begin{figure}[t!]
 \begin{center}
 \includegraphics[scale=0.45]{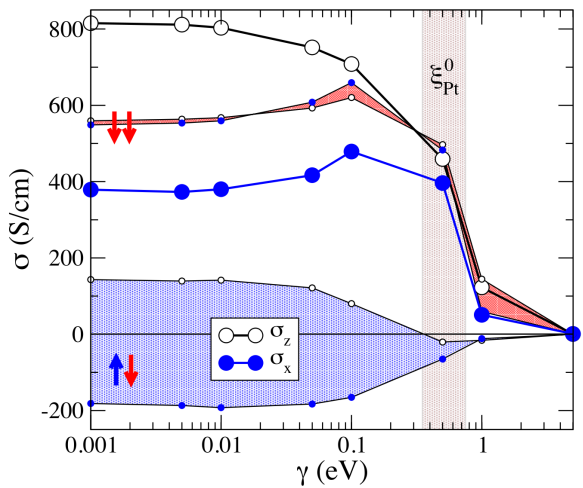}
 \end{center}
 \caption{ (color online)
   Dependence of the $\sigout$ (open circles) and $\sigin$ (filled
   circles) in FePt on quasiparticle damping rate $\gamma$, calculated
   within the Kubo-\v{S}treda formula. The area between $\uu$- and
   $\ud$-AHC for two different magnetization directions is shaded in 
   red and blue, respectively.}
 \end{figure}

Finally, we investigate the influence of disorder on the AHC and its
  anisotropy in FePt.  For this purpose, using the tight-binding 
  formulation in terms of the Wannier functions, we employ the Kubo-\v{S}treda
  formula~\cite{Streda} for the AHC in the "constant $\gamma$ 
  approximation", assuming that the quasiparticle damping rate $\gamma$
  is independent of the orbital~\cite{Tanaka}. The validity of this
  approximation for investigations of SHE and AHE in transition metals
  has been demonstrated~\cite{Tanaka,Kontani}. The results of our calculations
  for $\sigout$ (open circles) and $\sigin$ (filled circles) as a function of 
  disorder characterized by quasiparticle lifetime $1/\gamma$ are presented 
  in Fig.~3, and allow for a simple explanation in terms of energy 
  scales discussed previously.

Upon increasing disorder, isotropic $\sigma^{\uu}$ (red shaded area)
  stays practically constant until $\gamma$ of 0.1~eV, then decreases
  upon further increasing $\gamma$ and disappears at $\gamma$ of
  4$-$5~eV $-$ a value, which characterizes the decay of the
  $\uu$-cumulative AHC with energy, c.f. Fig.~2, and roughly corresponds
  to the characteristic band width in FePt.  The $\ud$-AHC (blue shaded
  area), on the other hand, decays much faster, and disappears at the
  value of $\gamma$ around 0.5~eV corresponding to the width of
  $A^{\ud}$ in Fig.~2. This value can be traced back to the Pt SOI
  strength $\xi_{\rm Pt}^0$ (grey shaded area in Fig.~3), which
  emphasizes Pt origin of spin-flip contribution to the AHC. Upon
  $\gamma$ reaching $\xi_{\rm Pt}^0$ the interband coherence necessary
  for a build-up of $\ud$-AHC is destroyed and $\sigma^{\ud}$ goes to
  zero. The relative robustness of $\sigma^{\uu}$ with respect to
  $\gamma$, as compared to $\sigma^{\ud}$, underines the fact 
  that the spin-flip transitions, living on a different energy scale, are 
  much more sensitive to the degree of crystallinity, and are affected
  stronger by disorder. Overall, after adding up $\uu$- and $\ud$-AHC
  we observe that while the total $\sigout$ (large open circles) monotonously
  decreases with $\gamma$ and drops significantly upon $\gamma$ 
  reaching $\xi_{\rm Pt}^0$, $\sigin$ (larde filled circles) stays relatively 
  constant in this range of disorder, with both conductivities vanishing at 
  $\gamma$ of several eV. This qualitatively different behavior of $\sigout$ 
  and $\sigin$ upon increasing disorder brings us to a conclusion that up to 
  a certain extent, the degree of disorder in FePt serves as the SOI strength
  $\xi_{\rm Pt}$, c.f.~Fig.~2, in accord to experimental findings~\cite{Chen}.

To conclude, we predict a strong anisotropy of the intrinsic AHC in FePt and 
  FePd alloys. We show, that while in FePt the AHC anisotropy arises purely 
  due to Pt-driven spin-flip transitions in a small energy window around $E_F$, 
  upon descreasing the SOI strength on Pt atoms, the sign of this anisotropy 
  and its nature can be changed in Pd-containing alloys. We also demonstrate 
  that in FePt the AHE comes from different types of atoms depending on the 
  direction of the magnetization and that the degree of disorder in the samples 
  of FePt can serve as an effective SOI strength of Pt atoms.

We acknowledge discussions with M. Le\v{z}ai\'c, Ph. Mavropoulos 
  and K. M. Seemann. We also thank HGF-YIG Programme VH-NG-513
  for funding and supercomputers JUROPA and JUGENE for computational time.


\begin{thebibliography}{99}

\bibitem{Nagaosa:review} N. Nagaosa {\it et al.}, Rev. Mod. Phys. {\bf 82}, 1539 (2009)
\bibitem{Hirsch} J.E. Hirsch, Phys. Rev. Lett. {\bf 83}, 1834 (1999)
\bibitem{Shi} J. Shi {\it et al.}, Phys. Rev. Lett. {\bf 96}, 76604 (2006)
\bibitem{Zutic} I. \v{Z}uti\'c {\it et al.}, Rev. Mod. Phys. {\bf 76}, 323 (2004)
\bibitem{MAE} C. Andersson {\it et al.}, Phys. Rev. Lett. {\bf 99}, 177207 (2007)
\bibitem{Kontani} T. Naito {\it et al.}, Phys. Rev. B {\bf 81}, 195111 (2010)
\bibitem{Gradhand} M. Gradhand {\it et al.}, Phys. Rev. Lett. {\bf 104}, 186403 (2010)
\bibitem{Seemann} K. Seemann {\it et al.}, Phys. Rev. Lett. {\bf 104}, 076402 (2010)
\bibitem{Yao} Y. Yao {\it et al.}, Phys. Rev. Lett. {\bf 92}, 037204 (2004)
\bibitem{Souza:2006} X. Wang {\it et al.}, Phys. Rev. B {\bf 74}, 195118 (2006)
\bibitem{wannier} A.~A.~Mostofi {\it et al.}, Comput. Phys. Commun. {\bf 178}, 685 (2008)
\bibitem{fleur} www.flapw.de
\bibitem{FLAPW-MLWFs} F. Freimuth {\it et al.}, Phys. Rev. B {\bf 78}, 035120 (2008)
\bibitem{footnote2} For $L1_0$ FePd we used the values of $c$ and $a$
                              of~7.15~a.u.~and~5.12~a.u., respectively.  
\bibitem{Souza:2009} E. Roman {\it et al.}, Phys. Rev. Lett. {\bf 103}, 097203 (2009)
\bibitem{Cooper} B. R. Cooper, Phys. Rev. {\bf 139}, A1504 (1965)
\bibitem{footnote1} In L1$_0$ FeNi alloy with $\xi_{\rm Ni}^0$ of 0.05~eV~our 
           calculated values of $\sigma_z$ and $\sigma_x$ constitute 165.7 and 284.8~S/cm,
           respectively, with $\ud$-contribution of about 30~S/cm. Corresponding spin-flip 
           anisotropy $\Delta\sigma^{\ud}$ is only 0.9~S/cm.   
\bibitem{Streda} P.~\v{S}treda, J. Phys. C {\bf 15}, L717 (1982)
\bibitem{Tanaka} T. Tanaka {\it et al.}, Phys. Rev. B {\bf 77}, 165117 (2008)
\bibitem{Chen} M. Chen {\it et al.}, arXiv: 1004.0548v3 (2010)

\end{thebibliography}
\end{document}